\documentclass[sigconf]{acmart}


\renewcommand\footnotetextcopyrightpermission[1]{}

\copyrightyear{2026}

\usepackage[utf8]{inputenc}
\usepackage{cleveref}
\usepackage{algorithm2e}
\usepackage{color}
\usepackage{framed}
\usepackage{xcolor}
\usepackage{enumitem}
\usepackage{natbib}
\usepackage{acmart-taps}

\usepackage{tikz}
\usetikzlibrary{arrows.meta,positioning}

\newcommand{\citesec}[1]{Section~\ref{sec:#1}}

\definecolor{rqanswerbg}{gray}{0.90}
\newenvironment{rqanswer}{%
  \setlength{\fboxsep}{8pt}%
  \setlength{\fboxrule}{0.8pt}%
  \colorlet{shadecolor}{rqanswerbg}%
  \begin{shaded}%
}{%
  \end{shaded}%
}

\usepackage{xspace}

\newsavebox{\promptbox}
\newenvironment{promptBox}
  {\begin{lrbox}{\promptbox}%
   \begin{minipage}{\columnwidth}
   \setlength{\parindent}{0pt}%
   \color{black}}
  {\end{minipage}%
   \end{lrbox}%
   \noindent\fcolorbox{gray!30}{gray!10}{\usebox{\promptbox}}}

\newcommand{\ApproachName}[0]{Segmentation-Tool-based approach}

\begin{document}
\title{\texorpdfstring{Imperfect Visual Verification for Code Edition :\\ A case study on TikZ}{Imperfect Visual Verification for Code Edition : A case study on TikZ}}

\author{Charly Reux}
\affiliation{%
  \institution{Univ Rennes, Inria, IRISA, INSA}
  \city{Rennes}
  \country{France}}
\email{charly.reux@inria.fr}

\author{Mathieu Acher}
\affiliation{%
  \institution{Univ Rennes, Inria, CNRS, IUF, IRISA}
  \city{Rennes}
  \country{France}}
\email{mathieu.acher@irisa.fr}
\date{March 2026}

\author{Djamel Eddine Khelladi}
\affiliation{%
  \institution{Univ Rennes, Inria, CNRS, IRISA}
  \city{Rennes}
  \country{France}}
\email{djamel-eddine.khelladi@irisa.fr}

\author{Clément Quinton}
\affiliation{%
  \institution{Univ. Lille, CNRS, Inria}
  \city{Lille}
  \country{France}}
\email{clement.quinton@univ-lille.fr}

\author{Olivier Barais}
\affiliation{%
  \institution{Univ. Rennes, IRISA, Inria}
  \city{Rennes}
  \country{France}}
\email{olivier.barais@irisa.fr}



\begin{abstract}
LLMs have significantly advanced code generation, enabling the synthesis of functional programs. While recent systems achieve strong performance on many coding benchmarks, tasks involving programs such as TikZ that generate visual artifacts remain challenging, in particular on visual code customization. 
Unlike generation from scratch, customization requires localized, semantics-preserving edits: the model must locate relevant code, modify it according to the instruction, and preserve the remaining structure and rendering.
 \textit{Approaches based on post-hoc iterative refinement/correction} where a verifier provides feedback to guide corrections, have shown promise.  
However, in the case of programs with a visual outcome such as in TikZ, where correctness is harder or likely impossible to formalize and evaluate automatically, deterministic verifiers do not exist. Hence, developers can only rely on imperfect verifiers. 

In this paper, we conduct an empirical study to answer: \emph{to what extent can iterative refinement remain effective when the verifier itself is unreliable?} 
We use TikZ as a focused case study that isolates the core difficulties of the problem (weak code structure, fine-grained visual semantics, and difficult feature localization) in a controlled and challenging setting.
We define visual code customization as an iterative editing problem with an imperfect oracle, and introduce a framework for analyzing such iterative refinements.
We conduct a large-scale study and evaluate multiple LLM-based and tool-augmented visual verifiers within iterative refinement pipelines, and perform extensive manual annotation of refinement trajectories to assess verifier behavior and feedback quality. 

Our findings show that even imperfect verifiers can determine with moderate accuracy whether visual instructions are applied to code, achieving F1-scores up to 0.815. Feedback improves iterative refinement, especially for weaker models, adding 11–20 perfect customizations for Qwen3-vl-30b-a3b-Instruct, while stronger models like Gemini-3 gain fewer improvements (+5) but benefit more from accurate verification that prevents premature acceptance. Feedback is effective only when it precisely identifies image issues, provides actionable guidance, addresses all relevant problems, and remains grounded in the original instruction.

\end{abstract}

\maketitle
\section{Introduction}

Large language models (LLMs) have significantly advanced automated code generation, enabling the synthesis of functional programs from natural language descriptions \cite{xu_systematic_2022}. While recent systems achieve strong performance on many coding benchmarks, a class of tasks remains challenging: programs that generate \textit{visual artifacts}~\cite{zhang_artifactsbench_2025,wu_plot2code_2024,lin_vcode_2025}, such as TikZ, SVG, or interactive web visualizations (e.g., D3.js).
In these settings, correctness is not solely determined by syntactic or functional properties of the code, but by high-level visual properties such as geometric structure, spatial relationships, and layout semantics. However, such properties are difficult to infer or verify from code alone.

A particularly challenging instance of this problem is \textit{visual code customization}, where an existing program must be modified according to a high-level visual instruction (e.g., \emph{"reduce the size of the green box on the left and put it parallel to the yellow one"}). 
Unlike generation from scratch, this customization requires localized, semantics-preserving edits: the model must identify where in the code a visual element is defined, modify it according to the instruction, and preserve the remaining structure and rendering.
This problem arises in a wide range of domains, including web interfaces, SVG graphics, scientific diagrams, games, and 3D modeling.
Although prior work on multimodal code generation (e.g., image-to-code or text-to-diagram synthesis) has shown promising results~\cite{rodriguez_starvector_2025,jiang_screencoder_2025,wei_words_2024}, recent studies indicate that LLMs still struggle to reliably perform such targeted visual modifications~\cite{reux_llm_2025}. Indeed,~\citeauthor{reux_llm_2025} showed that GPT-4o successfully completed only 26\% of TikZ customization tasks.

One promising paradigm for improving generation quality is \textit{post-hoc iterative refinement/correction} \cite{kamoi_when_2024,pan_automatically_2024}, where an LLM proposes candidate solutions and a verifier provides feedback to guide subsequent improvements (see \cref{dia:ite_refinement}).
This paradigm underlies a growing class of systems, including coding agents and Reinforcement Learning with Verifiable Rewards (RLVR), where feedback signals are typically deterministic, boolean, and highly reliable. 
Examples include unit test execution for code generation, constraint checking for structured outputs, or formal proof checkers in theorem-proving environments.
In such settings, refinement is effective because the verifier acts as a near-perfect oracle.


However, many real-world generation tasks lack such deterministic verification signals, making it challenging to design reliable refinement loops. When outputs are primarily visual, correctness is difficult (if not impossible) to formalize and evaluate automatically. 
For visual code such as TikZ, HTML/CSS, SVG, or SCAD, compilation or execution only ensures syntactic validity, not semantic correctness of the rendered image. 
As a result, refinement must rely on \textit{imperfect verifiers}, often implemented as LLM- or VLM-based evaluators assessing visual similarity or instruction compliance.
An imperfect verifier can be characterized along three dimensions: \emph{(i)} false positives, where incorrect outputs are mistakenly accepted, \emph{(ii)} false negatives, where correct or partially correct outputs are rejected, and \emph{(iii)} feedback quality, defined as the extent to which feedback is specific, actionable, and aligned with the underlying error. 
Such imperfections make the refinement process challenging: false positives lead to premature termination, while poor feedback prevents meaningful improvement.
This motivates the core research question of this work: \emph{To what extent can iterative refinement remain effective when the verifier itself is unreliable?}

This paper investigates this question through a focused and controlled case study on TikZ.
As detailed in Section~\ref{sec:background}, TikZ amplifies three core software engineering challenges while controlling away confounding factors such as multi-file architectures or framework-specific tooling: \emph{(i)} weakly structured and highly compositional code, which complicates visual element localization; \emph{(ii)} high sensitivity to small edits, where minor code changes produce subtle but critical visual differences; and \emph{(iii)} abstract geometric constructs, which are difficult for current vision-language models to interpret reliably.

To address this problem, 
we frame visual code customization as an iterative editing process under an imperfect oracle, and we introduce a framework for analyzing such iterative refinements.
Building on this framework, we conduct a large-scale empirical study using a refined version of the VTikZ dataset~\cite{reux_llm_2025}. 
In particular, we evaluate multiple LLM-based and tool-augmented visual verifiers within iterative refinement pipelines, and we perform extensive manual annotation of refinement trajectories to assess verifier behavior and feedback quality. We selected Qwen-vl-30b-a3b and Gemini-3-flash as the backbone models for our evaluation.

The results yield three key insights.
First, imperfect verifiers can correctly assess whether visual instructions are applied to code, achieving F1-scores up to 0.815.
Second, feedback significantly improves refinement for weaker models, yielding gains of +11 to +20 perfect customizations for Qwen3-vl-30b-a3b-Instruct, whereas stronger models (Gemini-3) benefit less from feedback and more from accurate classification that prevents premature acceptance. 
Third, feedback is effective only when it is actionable: it must precisely identify visual issues, provide concrete guidance, cover all relevant problems, and remain grounded in the original instruction.

To summarize, this work makes the following contributions:
\begin{itemize}
    \item We frame the general problem of iterative code customization with imperfect visual verifiers, TikZ being one challenging instance.
    \item We construct a large-scale dataset of manually annotated customization within iterative refinement trajectories.
    \item We provide the first empirical analysis of feedback on LLM refinement performance in the customization scenario.
    \item We derive a taxonomy of feedback behaviors and failure modes of LLM-based visual verifiers through systematic manual analysis of refinement trajectories. 
    \item We provide the dataset and a replication package of our experiments.
    
\end{itemize}

\section{Background and Motivation}
\label{sec:background}


This section introduces the necessary background and further motivates our work.

\paragraph{Problem formalization.} 
We formalize iterative code customization as the problem of refining a code artifact $c$ through a sequence of updates guided by an imperfect verifier. At each iteration, the generator proposes edits to the code, the verifier provides potentially noisy feedback and a score, and the process repeats until a stopping criterion or refinement budget is reached. The goal is to produce a final code artifact that best matches the intended visual outcome.

Let $c_0$ denote the initial code (a diagram).  
At iteration $t$, the generator produces the next code based on the current code, the verifier feedback, and the score:
\[
c_{t+1} = G(c_t, f_t),
\]
where $f_t = V(c_t)$ represents the feedback and score provided by the verifier.

For a complete refinement loop:
\[
c_0 \overset{G}{\to} c_1 \overset{V}{\to} f_1 \overset{G}{\to} \dots \overset{G}{\to} c_n.
\]
The quality of the final code $c_n$ is measured with respect to the intended visual goal $v^*$:
\[
Q(c_n, v^*),
\]
The iterative refinement protocol we formalize in \cref{sec:execution} provides a concrete operational instance of this process, enabling empirical evaluation with real LLM generators and verifiers.

\paragraph{TikZ as a testbed.}
 \begin{figure}
    \centering
    \includegraphics[width=\linewidth]{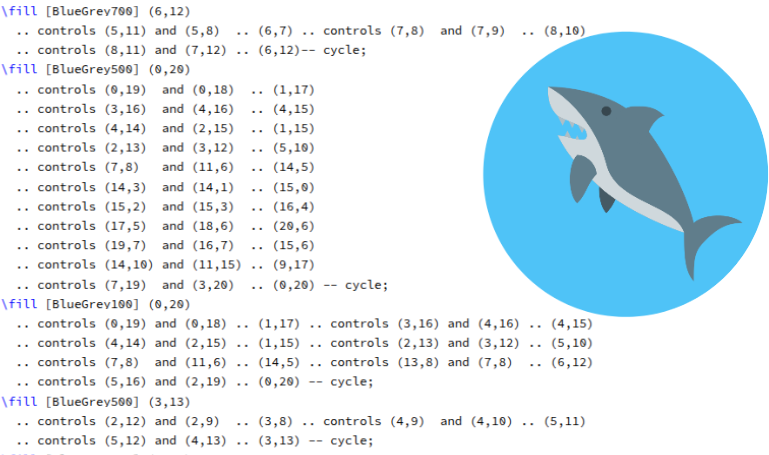}
   \caption{
       Excerpt of a TikZ code.
   }
    \label{dia:example_context}
      \Description{Excerpt of a TikZ code that creates a shark drawing}
\end{figure}
The following work focuses on the \textit{TikZ} language, a \LaTeX Package that allows the creation of graphic elements, mainly for scientific diagrams, though it can also be used for drawings and cartoon characters. TikZ programs define visual elements such as shapes, edges, and layouts through code, making them a representative example of programs that generate visual artifacts. 

We use TikZ as a \textit{focused and controlled case study} for iterative refinement with imperfect verifiers. Unlike broader settings such as web development, which involve multi-file structures, modular architectures, and mature visual tooling, TikZ controls away these confounding factors while \textit{amplifying} the challenges central to our study: the code is weakly structured and lacks modular boundaries, making feature localization difficult; visual outputs require fine-grained geometric reasoning; and small code changes can produce subtle yet critical visual differences. The code is often sufficiently opaque that even advanced LLMs fail to interpret what it represents. For instance, given the code of \Cref{dia:example_context}, GPT-4 misinterprets it as a ``stylized abstract landscape'' and Gemini-3 as a seagull, when it in fact represents a shark.
TikZ thus provides an idealized yet practically relevant setting that isolates the mechanism of interest (imperfect visual verification and fine-grained code localization), providing analytical leverage for studying verifier--generator dynamics. Customizing TikZ code from high-level visual instructions requires translating visual-level requests into localized code changes while preserving the overall diagram structure.



\paragraph{Iterative refinement approach.}
 
 \begin{figure}
    \centering
    \includegraphics[width=\linewidth]{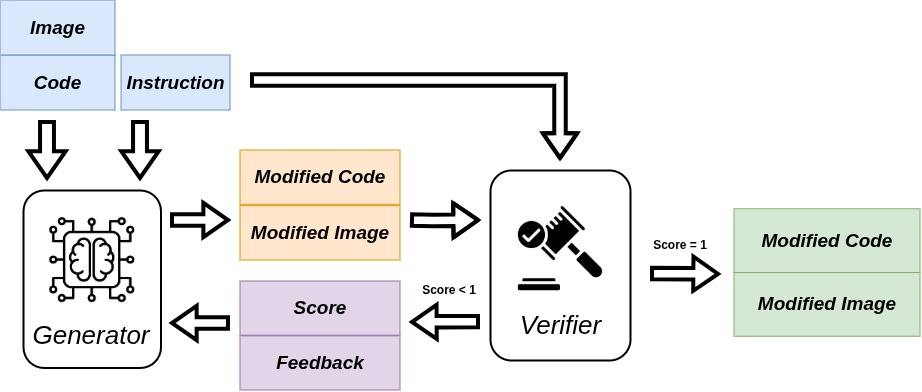}
    
    \caption{Post-hoc Iterative Refinement.}
    \label{dia:ite_refinement}
    \Description{Overview of the Post-hoc Iterative Refinement in the context of TikZ code customization}
    
\end{figure}

To address this challenge, we explore \textit{post-hoc iterative refinement}, illustrated in \Cref{dia:ite_refinement}. In this paradigm, a \texttt{Generator} attempts to modify the code according to the visual instruction, while a \texttt{Verifier} evaluates whether the instruction has been correctly applied, given the code and the rendered image. The verifier returns a score and textual feedback, which can be used to iteratively refine the generated solution.

For example, consider a simple customization such as removing a specific feature in the TikZ drawing in \Cref{dia:example_context}. A user instruction might ask that the teeth of the shark be removed. The Generator may mistakenly identify the teeth and instead remove the fins. The verifier can detect this discrepancy and guide the generator toward the correct edit in subsequent iterations. Alternatively, the verifier may only deteriorate the code customization if incorrect feedback is further returned. 

\paragraph{Challenges and research gap.}

While iterative refinement has shown promising results in domains such as code optimization, readability improvement, constrained generation, and mathematical reasoning~\cite{madaan_self-refine_2023}. To the best of our knowledge, its effectiveness in visual customization settings remains unclear. In our case, verifiers are inherently imperfect, as visual instructions may admit multiple valid solutions and cannot be evaluated through deterministic tests. As a result, it is uncertain whether LLM-based verifiers can reliably detect whether a visual instruction has been correctly applied, and whether their feedback can effectively guide refinement. Our work fills this gap. 




\section{Studied Verifiers}
\label{sec:baseline_verifiers}

The verifiers we evaluate fall into two categories: (1) \textit{basic verifiers}, which are variations of multimodal LLM-based verifiers where either the input modality is changed (\textit{Text}, \textit{Visual}, or \textit{Text+Visual}) or the prompting technique is modified (\textit{Property}); and (2) \textit{complex verifiers}, which follow a more elaborate code-based approach. All system prompts and settings are available in the Git repository\footnote{\textcolor{blue}{\url{https://github.com/CharlyReux/verifier_problem}}}
Each verifier outputs a numeric score and textual feedback.  
For consistency, we map all scores to the interval $[0,1]$.  
For most verifiers, during prompting, we ask models to produce an integer score from 0 to 100, then divide by 100, as manual testing showed that models are more likely to assign a perfect score when the range is $[0,1]$ rather than $[0,100]$. 
We also define specific scoring thresholds (0–49, 50–79, 80–99, 100), depending on the extent to which the customization is applied.

\subsection{Simple LLM-based Verifier}

These verifiers differ from each other in the modality provided to the multimodal model: 1) \textit{Text}, where the initial and customized code are given; 2) \textit{Visual}, where 
the initial and customized images are fed to the model; and 3) \textit{Text+Visual}, where both modalities are provided (images and codes). 

\subsection{Property-based Verifier}

This verifier follows a two-step procedure. 
First, given the instruction and the original image, it generates a list of properties that the customized image should satisfy. Then, each property is evaluated independently by prompting the same model in a new context to determine whether the property holds, using a prompt similar to that of the simple LLM-based verifiers. 
The final score is the average score of the per-property scores.

\subsection{Agentic Vision Verifier}
\label{base:agentic_vision}

This verifier is based on recent work by Google, which leverages code execution to ground the LLM’s reasoning\footnote{\color{blue}{\url{https://blog.google/innovation-and-ai/technology/developers-tools/agentic-vision-gemini-3-flash/}}}. 
The approach generates Python code to manipulate and/or analyze an image in order to extract more fine-grained information. 
It is analogous to visual programming and agentic approaches, and serves as a solid code-based baseline.

\subsection{Segmentation and Tool-based Verifier}
\label{sec:base:segtool}

We implemented an approach that leverages the code-generation capabilities of LLMs to produce Python code. By prompting the model with a precise specification, we ensure that the generated code uses a set of oracle (functions). Each oracle is designed to perform a specific visual verification on a change between two images (e.g., color, shape, position).

\subsubsection{Code Generation and Score Computation}

The first step of this approach is to prompt the model to generate a verifier function that uses the available oracles.  This function is then executed on the customized image and returns a score between 0 and 1.

For example, to verify that a circle is changed to blue and its width is doubled, the \textit{color} and \textit{size} oracles can be combined using the \textbf{and} operator.
\begin{verbatim}
def test_valid_customization() -> bool:
 return color("circle", "blue") 
    and size("circle", (2, 1))
\end{verbatim}

Each oracle internally produces a probability, which are combined to compute a final score, computed using probabilistic Boolean combination.

\subsubsection{Automatic Feedback Generation}

 Each oracle produces automatic feedback describing the detected discrepancy. When the generated verifier code combines multiple oracles using logical operators, feedback is reported only when the corresponding combined probability exceeds a predefined threshold of 0.9.


\subsubsection{Oracle Catalog}

Our oracle catalog includes checks for spatial relations (placement, position, alignment, mirrored), visual attributes (color, size, shape), containment and quantity (within, present, count), rotation (angle), and complex visual properties (visual\_property). Each oracle leverages Gemini-2.5 segmentation to identify visual elements from textual descriptions.
With this type of method, completeness for an oracle catalog is difficult to achieve, and there is a risk of not covering every possible use case. This limitation is inherent to the catalog-based approach.

\section{Evaluation Methodology}
\label{sec:eval_methodo}

\begin{figure}
    \centering
    \includegraphics[width=\linewidth]{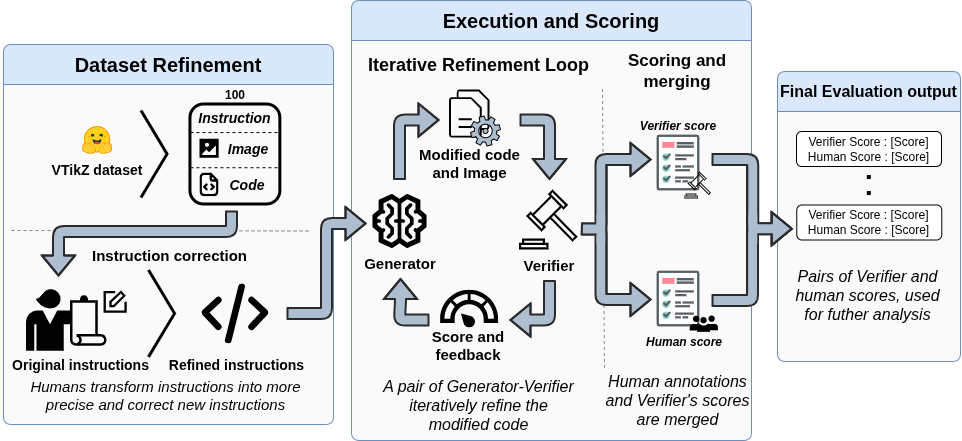}
   \caption{
       Complete evaluation protocol.
   }
    \label{diag:eval_protocol}
    \Description{Overview the complete evaluation protocol used in the study}
\end{figure}

This section presents our evaluation methodology, summarized in \cref{diag:eval_protocol}. We first refine the VTikZ dataset by correcting and improving the existing instructions. Next, we run the iterative post-hoc refinement loop using selected verifier/generator pairs. At each iteration, we collect the generated image (produced from the customized code) along with the score assigned by the verifier. The resulting images are then evaluated by human annotators, who are provided with the corresponding instruction and the original image, and asked to rate them on a 1–5 scale. This yields pairs of human and verifier scores, which form the basis of our subsequent analysis.
We analyze verifier performance along two dimensions: scoring accuracy and feedback quality.
Based on this distinction, we formulate the following research questions:

\begin{enumerate}[label=\textbf{RQ\arabic{enumi}}]
    \item \textit{To what extent do verifiers correctly assess the application of a visual instruction to code?}
    \begin{enumerate}[label=RQ \arabic{enumi}.\arabic*]
        \item Can they correctly detect whether a customization has been applied or not?
        \item Do their scores correlate with human judgments?
    \end{enumerate}

    \item \textit{How does the feedback created and the score assigned by verifiers relate to performance increases in the customization task?}
    \begin{enumerate}[label=RQ \arabic{enumi}.\arabic*]
        \item Does the quality of feedback matter?
        \item To what extent does the score matter to achieve better solutions?
    \end{enumerate}
   \item \textit{How does feedback influence the interaction between the generator and the verifier?}
    \begin{enumerate}[label=RQ \arabic{enumi}.\arabic*]
        \item Which types of feedback lead to errors, and which guide the generator toward better solutions?
        \item How do generator LLMs interpret and act upon the feedback they receive?
        \item What makes a Generator create a perfect solution?
    \end{enumerate}
\end{enumerate}

\subsection{Dataset}

We base our evaluation on the existing \textit{VTikZ} dataset~\cite{reux_llm_2025}. 
We performed a manual pass to correct typographical errors, resolve ambiguities, and improve clarity.
Every instruction specifies a visual modification to be applied to an existing TikZ program.
The corresponding original code and rendered image serve as the starting point for the iterative refinement process.

\subsection{Execution}
\label{sec:execution}

We apply the iterative refinement to each customization. At every iteration, the generated code, rendered image, and verifier's score are recorded. 
This results in a sequence of refinement steps for each customization, which are later used for human annotation and quantitative analysis.

\textit{Formalization of the Iterative Refinement Protocol.}
For a given customization, we consider a Generator $G$ and a Verifier $V$.
The Generator $G$ exposes two operations: $\textit{generate}$ and $\textit{adjust}$. The $\textit{generate}$ operation takes a code and an instruction as input and produces a modified code. The $\textit{adjust}$ operation takes feedback and a score and returns an updated code.
The Verifier $V$ takes an instruction and a context as input, with the context containing the original and customized rendered images and/or their corresponding source codes. The verifier returns a score between 0 and 1 along with textual feedback.

\begin{algorithm}
\LinesNumbered
\caption{Iterative Refinement Execution}
\label{algo_itrefine}
\SetKwFunction{Generate}{G.generate}
\SetKwFunction{Adjust}{G.adjust}
\SetKwFunction{Validate}{V}
\SetKwFunction{Renderer}{Renderer}
\SetKwInOut{Input}{Input}
\SetKwInOut{Output}{Output}

\Input{Code $c$ and instruction $i$}
\Output{List of $(c', r', s)$}

\BlankLine
$s \gets 0$; $\tau \gets 1$; $T \gets 5$\;
$r \gets \Renderer{c}$\;
$c' \gets \Generate{c,i}$\;\label{line:generate}
$D \gets []$\;

\For{$j \gets 0$ \KwTo $T-1$}{\label{line:loop}
    $r' \gets \Renderer{c'}$\;
    $\textit{context} \gets (r,r',c,c')$\;\label{line:context}
    $(f, s) \gets \Validate{i,\textit{context}}$\;\label{line:verifier}
    $D.\text{append}(c',r',s)$\;\label{line:save_d}
    \If{$s \ge \tau$}{
        \Return{$D$}\label{line:stop_thresh}
    }
    
    $c' \gets \Adjust{f,s}$\;\label{line:adjust}
}

\Return{$D$}
\end{algorithm}

The process for a pair $(G, V)$ is outlined in Algorithm~\ref{algo_itrefine}. Given an initial code and instruction, the generator $G$ produces a customized code $c'$ (l.\ref{line:generate}).
An iterative refinement loop is then performed (l.\ref{line:loop}). At each iteration, a context is constructed (l.\ref{line:context}) that includes the initial and generated images and codes $(r, r', c, c')$. This context is provided to the verifier $V$ (l.\ref{line:verifier}), which outputs feedback and a score $(f, s)$.
If the score satisfies a predefined threshold (set to 1 in our experiments), the process terminates (l.\ref{line:stop_thresh}). Otherwise, $G$ uses $(f, s)$ to produce an updated version of the code (l.\ref{line:adjust}). The loop continues until the threshold is met or the maximum number of iterations $T$ is reached.
At each iteration, the current code and image $(c', r')$, together with the score $s$, are recorded (l.\ref{line:save_d}) and returned upon termination.
This process is executed on all the customizations in the dataset, and returns code-image-score triplets.

\subsection{Manual Annotation}

The annotation process is conducted through a web interface in which annotators evaluate individual refinement steps rather than only the final output. Upon connection, each annotator is shown an image randomly sampled from all code-image-score triplets. The corresponding original image and customization instruction are also displayed. The annotator is then asked to rate how well the instruction was applied on an ordinal scale from 1 to 5.

\begin{itemize}
    \item 1 - Not applied at all
    \item 2 - Slightly applied
    \item 3 - Moderately applied
    \item 4 - Mostly applied
    \item 5 - Perfectly applied
\end{itemize}
This gives us, for each step, triplets of \textit{image}, \textit{verifier score}, and \textit{human score}.

Eight annotators participated in the manual annotation process. Although the dataset contains many generated images, several correspond to identical or very similar code modifications produced by the LLMs. As a result, the annotations cover 3907 distinct items, which correspond to a larger total of 15345 generated instances in the dataset. This deduplication significantly reduced the amount of manual annotation required while still providing coverage of a larger set of generated outputs. More details about the human-annotated data are provided in \Cref{tab:annotation_stats}.
We measure inter-annotator agreement using Krippendorff’s $\alpha$, obtaining a value of 0.898, indicating strong agreement and supporting the reliability of the annotations.

\subsection{Evaluation Configurations}

Because we evaluate \textit{verification methods} rather than the \textit{models} themselves, we conduct experiments using two models: \textit{Qwen3-vl-30b-a3b-instruct} and \textit{Gemini-3-flash-preview}, with the Gemini model configured with minimal thinking effort. These models were chosen for their strong coding and multimodal capabilities.
 Each model is used both as a generator and as a verifier. Unless specified otherwise, all models are run with a temperature of 0.5.

We first evaluate the four \textit{basic verifiers} (\textit{Simple} and \textit{Property}) using Gemini-3 as the verifier model, paired with both Qwen and Gemini as generators. We then repeat the same experiment using Qwen as both the verifier and generator model.
In addition to these configurations, we evaluate the two \textit{complex verifiers} (Agentic Vision and Segmentation Tool-Based) paired with both Qwen and Gemini. Finally, we introduce two additional verifiers: (1) \textit{SentenceVerifier}, a trivial verifier that always returns the message "The customization is not applied, please apply it.", and (2) \textit{PreciseVerifier}, a CLIP-based verifier, executed last, which only returns a score measured using the maximum CLIP similarity between the customized image and the set of images annotated as perfect. Both of these new baselines allow us to assess the usefulness of the feedback and the score, which we interpret in \Cref{sec:rq1,sec:rq2}

Overall, this results in 20 generator-verifier pairs. Each pair is evaluated on the 100 TikZ customizations, with three repeated runs per customization and up to five refinement iterations per run.

\begin{table}[ht]
\centering
\caption{Human Annotation Statistics and Inter-Annotator Agreement}
\label{tab:annotation_stats}
\small
\resizebox{0.85\linewidth}{!}{
\begin{tabular}{ll}
\toprule
\#Annotators & 8 \\
Total number of distinct items & 3907 \\
Total number of items & 15345 \\
Total number of distinct annotated items & 4289 \\
Total number of annotated items & 17100 \\
Avg. annotations per Item & 1.11 \\
Avg. distinct \#items per annotator & 536 \\
Avg. \#items per annotator & 2138 \\
Krippendorff's $\alpha$ & 0.898 \\
\bottomrule
\end{tabular}

}
\end{table}

\section{Evaluation Results}
\label{sec:eval_results}

This section now presents our observed results answering our RQs.  


\subsection{\emph{RQ1:} Assessment of Visual Instruction Application on Code}
\label{sec:rq1}

\begin{figure*}
    \centering
    \includegraphics[width=\linewidth]{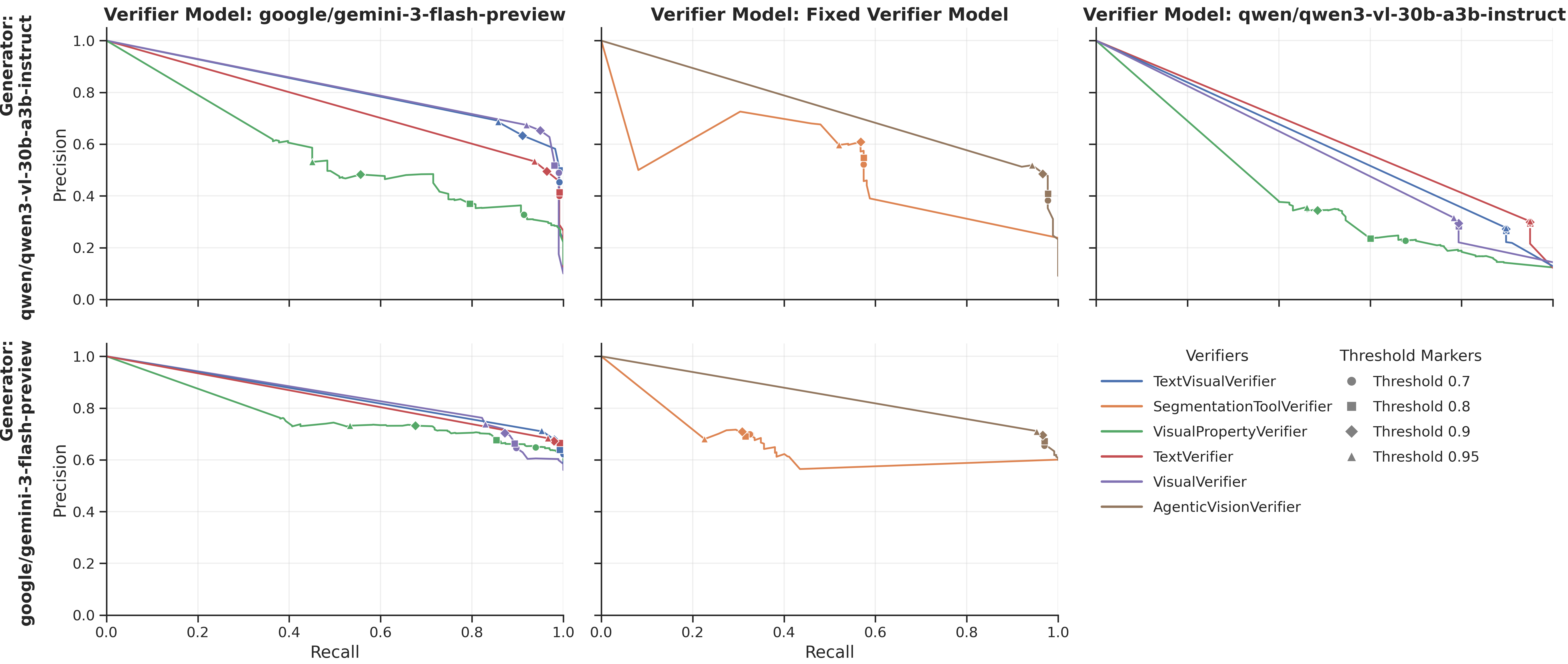}
   \caption{
       PR curves for each verifier.
   }
    \label{figure:pr_curves}
    \Description{Visualization of the Precision-Recall curves for all verifiers}
\end{figure*}
For this evaluation, we first focus on the ability of verifiers to identify perfect customizations. We therefore cast the task as a binary classification problem, where items rated 5 (“Perfectly applied”) by human annotators are treated as positive instances, and all others as negative. We then evaluate how well verifier scores can predict these positives using F1-score.
Figure~\ref{figure:pr_curves} presents the precision-recall (PR) curves for all evaluated generator-verifier pairs. Overall, \textit{simple LLM-based verifiers} (Text, Text+Visual, Text-only) as well as AgenticVision tend to assign high scores to most candidate customizations. As a result, their PR curves are concentrated in the high recall region, 
with no operating point (except near 0) achieving both high precision and low recall.
This indicates a strong bias toward accepting outputs, yielding consistently high recall but limited precision.

Introducing more structured verification partially mitigates this behavior.
The VisualPropertyVerifier reduces the frequency of uniformly maximal scores, leading to a more balanced score distribution. 
However, this comes at the cost of overall effectiveness, with both precision and recall decreasing relative to simple LLM-based approaches. A similar trade-off is observed for the Segmentation and Tool-based approach, which performs similarly to VisualPropertyVerifier but yields overall underwhelming results.

\begin{table*}[t] 
\centering
\caption{Precision/Recall/F1 Scores for the best thresholds for llm verifiers}
\label{tab:best_F1_llms}
\resizebox{\linewidth}{!}{
\begin{tabular}{ll|rrrr|rrrr|rrrr|rrrr}
\toprule
Verifier  & Generator & \multicolumn{4}{c}{Text} & \multicolumn{4}{c}{Text+Visual} & \multicolumn{4}{c}{Visual} & \multicolumn{4}{c}{Visual Property} \\
 Model &  & Threshold & Precision & Recall & F1 & Threshold & Precision & Recall & F1 & Threshold & Precision & Recall & F1 & Threshold & Precision & Recall & F1 \\
\midrule
Gem. & Gem. & 1.000 & 0.683 & 0.967 & 0.801 & 0.980 & 0.712 & 0.952 & \textbf{0.815} & 1.000 & 0.762 & 0.822 & 0.791 & 0.667 & 0.645 & 0.970 & 0.775 \\
Gem. & Qwen & 1.000 & 0.533 & 0.937 & 0.680 & 1.000 & 0.691 & 0.857 & 0.765 & 1.000 & 0.674 & 0.919 & \textbf{0.778} & 0.875 & 0.484 & 0.715 & 0.578 \\
\midrule
Qwen & Qwen & 1.000 & 0.302 & 0.950 & \textbf{0.458} & 1.000 & 0.277 & 0.897 & 0.423 & 1.000 & 0.316 & 0.783 & 0.450 & 0.880 & 0.348 & 0.531 & 0.421 \\
\bottomrule
\end{tabular}
}
\end{table*}
\begin{table}[t] 
\centering
\caption{Precision/Recall/F1 Scores for the best thresholds for fixed verifiers}
\label{tab:best_F1_fixed}
\resizebox{\linewidth}{!}{
\begin{tabular}{l|rrrr|rrrr}
\toprule
Generator & \multicolumn{4}{c}{Agentic Vision} & \multicolumn{4}{c}{Segmentation-Tool Verifier} \\
 & Threshold & Precision & Recall & F1 & Threshold & Precision & Recall & F1 \\
\midrule
Gem. & 1.000 & 0.712 & 0.949 & \textbf{0.814} & 0.000 & 0.601 & 1.000 & 0.751 \\
Qwen & 0.980 & 0.519 & 0.943 & \textbf{0.669} & 0.920 & 0.609 & 0.568 & 0.587 \\
\bottomrule
\end{tabular}

}
\end{table}
To further quantify these trends, the best achievable F1-score for each approach is computed over all decision thresholds.
The corresponding precision and recall values are reported in \Cref{tab:best_F1_llms} and \Cref{tab:best_F1_fixed}. For readability, we abbreviate model names as follows: \textit{Gem.} (Gemini-3-flash-preview) and \textit{Qwen} (Qwen3-vl-30b-a3b-instruct).
For \textit{simple LLM-based Verifiers}, optimal F1-scores are consistently obtained at thresholds close to 1, confirming the tendency of these models to assign high scores, leading to high recall but relatively low precision.

Across all approaches, performance improves when paired with stronger generator models, a trend observed for both LLM-based verifiers (\Cref{tab:best_F1_llms}) and fixed approaches (\Cref{tab:best_F1_fixed}). 
Notably, AgenticVision achieves the highest overall F1-score when paired with Gemini (0.814), but not with Qwen (0.669).


\begin{table}[t] 
\centering
\caption{Kendall correlations for LLM and Fixed verifiers}
\label{tab:kendal_corrs_llm_fixed}
\resizebox{\linewidth}{!}{
\begin{tabular}{ll|rrrr}
\toprule
Verifier Model & Generator & Text & Text+Visual & Visual & Visual Property \\
\midrule
Gem. & Gem. & 0.172 & 0.448 & \textbf{0.456} & 0.287 \\
Gem. & Qwen & 0.718 & \textbf{0.765} & 0.709 & 0.530 \\
\midrule
Qwen & Qwen & \textbf{0.620} & 0.616 & 0.487 & 0.364 \\
\bottomrule
\end{tabular}
}
\\~\\
\resizebox{\linewidth}{!}{
\begin{tabular}{l|rrr}
\toprule
Generator & Agentic Vision & Precise Score & Segmentation-Tool Verifier \\
\midrule
Gem. & 0.441 & \textbf{0.534} & 0.003 \\
Qwen & \textbf{0.718} & 0.373 & 0.347 \\
\bottomrule
\end{tabular}

}
\end{table}


\Cref{tab:kendal_corrs_llm_fixed} reports Kendall correlation coefficients between verifier scores and all human judgments. In contrast to F1-scores, an opposite trend can be observed: correlation is higher when the generator model is weaker, suggesting that stronger generators produce outputs that are harder to discriminate.
Input modality also plays a key role.
When paired with weaker generators, combining textual and visual inputs (\textit{Text+Visual}) yields the strongest alignment with human judgments (correlation up to 0.765)
However, with stronger generators, relying solely on visual input (\textit{Visual}) leads to better alignment, indicating that textual signals may introduce noise once baseline output quality is high.


\begin{rqanswer}
    \textbf{RQ1:} 
    Among the evaluated approaches, only Gemini-3-Flash is reliable at detecting whether a customization has been applied, achieving F1-scores up to 0.815. However, it tends to overestimate correctness, resulting in high recall but lower precision. Correlation with human judgments is moderate when paired with Gemini as the generator (0.456), but substantially higher with Qwen (0.765).
    
    More complex verification strategies—such as code generation (AgenticVision, Segmentation-Tool-based) or decomposition into multiple verification steps (VisualProperty) - do not improve performance and can even degrade both accuracy and alignment with human judgments.
    
    Overall, for TikZ code, LLM-based verifiers (particularly Gemini-3-Flash) can be used to assess whether a visual instruction has been applied. However, this comes with a significant rate of false positives, reflecting a systematic bias toward high recall at the expense of precision.

\end{rqanswer}

\subsection{\emph{RQ2:} Overall Performances}
\label{sec:rq2}

\definecolor{PastelRed}{RGB}{244, 180, 180}
\definecolor{Lavender}{RGB}{200, 180, 255}
\definecolor{PastelPink}{RGB}{255, 200, 220}
\definecolor{PastelGreen}{RGB}{150, 200, 160}
\definecolor{PastelBlue}{RGB}{180, 210, 240}
\definecolor{Peach}{RGB}{255, 210, 170}
\definecolor{LightBrown}{RGB}{196, 134, 62}
\definecolor{LightGray}{RGB}{83, 83, 83}
\begin{figure*}
    \centering
    \includegraphics[width=\linewidth]{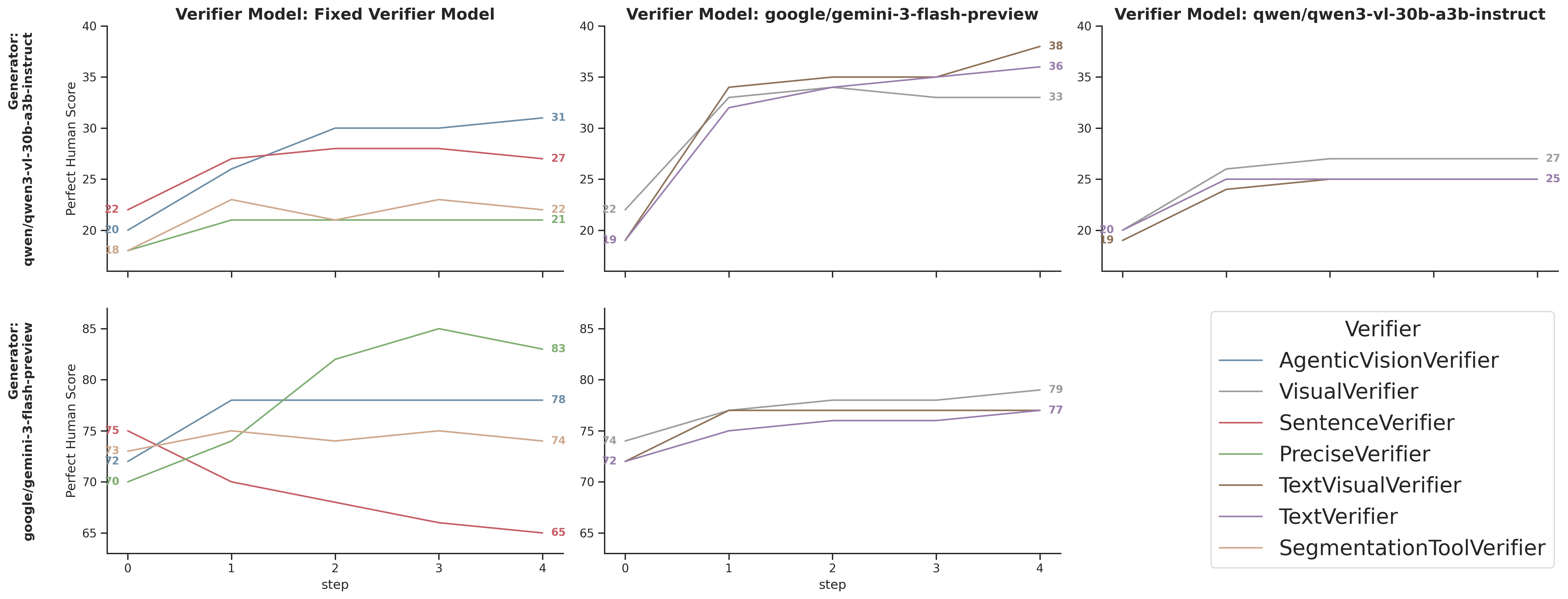}
   \caption{
       Total number of perfect customizations per step per verifier.
   }
    \label{figure:total_perfect_score}
    \Description{Visualization of the score reached at each step for each configuration evaluated}
\end{figure*}

\Cref{figure:total_perfect_score} reports the step-wise performance of all approaches, measured as the number of customizations that achieve a perfect score according to human annotators.

For Qwen as the \textbf{generator}, performance improves when paired with Gemini as the verifier, with gains of up to +16 and +17 of perfect customizations. 
The largest gains occur when Gemini has access to the generated code (\textcolor{Lavender}{TextVerifier}, \textcolor{LightBrown}{TextVisualVerifier}), although gains remain noticeable with visual-only input (\textcolor{LightGray}{VisualVerifier}). 
Compared to the baseline where Qwen acts as both generator and verifier, using Gemini leads to consistent gains, although they are more limited in the visual-only setting (+11). 
Pairing Qwen with the \textcolor{PastelRed}{SentenceVerifier}, which provides only coarse binary feedback (“customization not applied”), yields improvements comparable to being paired with itself, suggesting that minimal feedback can be useful for weaker models.

Improvements are more limited when Gemini is used as the \textbf{generator}. When paired with itself, the \textcolor{LightGray}{VisualVerifier} yields the best results, but only marginally (+2). Unlike Qwen, performance decreases when using the \textcolor{PastelRed}{SentenceVerifier}. This difference can be explained by the higher initial performance of Gemini (around 75 correct customizations versus 20–24 for Qwen), 
which reduces the need for additional exploration and increases the risk of degrading already correct outputs during refinement.

More complex verification strategies, such as \textcolor{PastelBlue}{AgenticVision} and the \textcolor{Peach}{Segmentation-Tool-based approach}, yield limited gains across both generators. 
Their performance remains comparable to, or slightly below, that of 
simpler LLM-based verifiers, indicating that added architectural complexity does not translate into improved refinement.

The \textcolor{PastelGreen}{PreciseVerifier} yields the largest performance improvements, reaching up to 85 perfect customizations at step 4. By leveraging similarity scores with known perfect outputs, it effectively halts successful trajectories early while allowing further exploration on unresolved cases. However, Qwen benefits less from this mechanism, with only marginal improvements (+2).

\begin{rqanswer}
    \textbf{RQ2:} 
    Feedback can improve refinement performance, but its impact strongly depends on the strength of the generator. For weaker models such as Qwen, informative feedback from stronger verifiers (e.g., Gemini-based) leads to substantial gains (33–36 perfect customizations), while coarse or indirect signals (e.g., \textcolor{PastelGreen}{PreciseVerifier}, \textcolor{PastelRed}{SentenceVerifier}) result in lower performance (28–30).

    For stronger models such as Gemini, the role of feedback differs: coarse feedback degrades performance, whereas more informative verifiers (\textcolor{LightBrown}{TextVisualVerifier}, \textcolor{LightGray}{VisualVerifier}) yield modest improvements. The best results are obtained with \textcolor{PastelGreen}{PreciseVerifier}, indicating that, at higher performance levels, accurate scoring and early stopping become more critical than exploratory feedback.
    
    Overall, feedback is beneficial for improving performance in the TikZ customization task, especially for weaker models. However, for stronger models, the precision and reliability of the verifier play a more critical role than the feedback.
    
\end{rqanswer}

\subsection{\emph{RQ3:} Qualitative Feedback analysis}
\label{sec:rq3}


To better understand LLM behavior during iterative refinement, we conduct a manual analysis of verifier-generated feedback. We focus on feedback produced by \textit{TextVisualVerifier}, \textit{PropertyVerifier}, and \textit{AgenticVisionVerifier}, as these approaches generate open-ended, visually grounded, natural language feedback. In contrast, other approaches either constrain feedback to predefined properties (\textit{VisualPropertyVerifier}) or rely on fixed oracle responses (\textit{\ApproachName}).

\begin{figure*}
    \centering
    \includegraphics[width=0.8\linewidth]{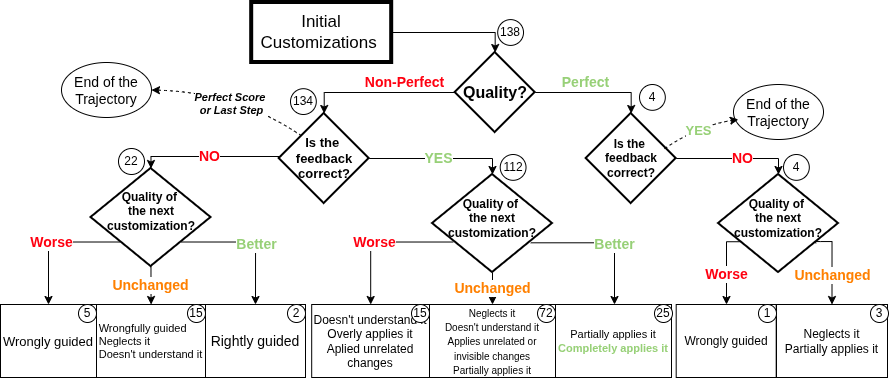}
   \caption{
       Observed Customization Trajectories, and subsequent interpretation/actions by the generator.
   }
    \label{fig:trajectory_visualization}
    \Description{Decision tree describing the customization Trajectories studied}
\end{figure*}

We construct a characterization of feedback in customization trajectory steps. We consider three dimensions:
\begin{itemize}
    \item (i) the quality of the initial customization (perfect vs.\ non-perfect)
    \item (ii) the correctness of the feedback (correct vs.\ erroneous with respect to the instruction or image)
    \item (iii) the quality of the subsequent customization (Worse, Unchanged, or Better)
\end{itemize}
For each output of a trajectory step, we analyze the feedback in terms of \textbf{aspects} (i.e., what the feedback contains) and \textbf{interpretations} (i.e., how it is acted upon by the generator).
The analysis is conducted over \textbf{138 randomly sampled tuples} of the form $\langle$initial customization, feedback, refined customization$\rangle$. Sampling was continued until \textbf{no new feedback patterns or trajectory behaviors were observed}, suggesting that the main trajectory patterns are adequately represented.
This enables us to identify recurring feedback patterns associated with improvement, stagnation, or degradation in customization quality.

We summarize this study in \cref{fig:trajectory_visualization}, which depicts a decision tree over customization trajectories. Starting from an initial customization, trajectories branch on the observed quality (Perfect vs. Non-Perfect) and the correctness of the feedback, leading to subsequent customization steps. At each decision node, we report the frequency of transitions, and at the leaves, we detail the generators’ interpretations and resulting actions (e.g., rightly guided, partially applied, neglected, or wrongly guided).

For incorrect feedback, the dominant failure modes are \textit{hallucinations} on the visual content or instruction and \textit{misunderstandings} of the instruction. 

For correct feedback, we identify several aspects, specifically, it may be \textbf{imprecise}, \textbf{non-guiding}, \textbf{incomplete}, or \textbf{forgetful}, which can lead to partial, incorrect, or missing application in the next customization.

We define these aspects as follows:
\begin{itemize}
    \item \textbf{Precision}: The feedback localizes and specifies issues clearly, rather than describing them in vague/abstract terms
    \item \textbf{Guidance}: The feedback goes beyond identifying errors and provides actionable directions for correction.
    \item \textbf{Completeness}: The feedback captures all necessary issues required to correctly refine the customization.
    \item \textbf{Forgetfulness}: The feedback "forgets" about the initial instruction, and only comments on the flaws of the diagram.
\end{itemize}

Independent of correctness, feedback may be interpreted and acted upon in different ways by the generator as it may be understood or not, and either applied, partly applied, or ignored.
We also observe recurring behaviors such as \textbf{invisible changes}, \textbf{stubbornness}, or \textbf{neglect} where the generator fails to modify the code despite receiving feedback. This may stem either from an inability to operationalize the feedback or from overconfidence in the current solution.

\textbf{Invisible changes} are generally reformatting, adding comments, or shifting the entire picture. Together with \textbf{neglect}, these are the main reasons why the majority of the customizations happening exhibit the same quality as the previous customizations, and are mainly due to Qwen's inability to correctly apply the feedback.


Finally, we select the behaviors that led to perfect customization. For these we have observed most/all of the aspect described above: the feedback is leverage-able and states what must be modified (\textbf{guiding}), it explicitly describes the what is right and what is wrong w.r.t the instruction (\textbf{not forgetful}) and it provides detailed information (\textbf{Precise}), and finally, it states all the points which are to be fixed (\textbf{Complete}).

\begin{rqanswer}
    \textbf{RQ3:} Feedback that leads to errors is either (i) incorrect, due to \textbf{hallucinations} or \textbf{misunderstandings}, or (ii) correct but ineffective, due to being \textbf{imprecise}, \textbf{non-guiding} or \textbf{forgetful}.
    Generator LLMs exhibit diverse interpretation behaviors, ranging from correct understanding and application to failure modes such as misunderstanding, partial application, \textbf{neglect}, and \textbf{stubbornness}.
    Given that the generator LLM can operationalize the feedback, our study shows that perfect customization is achieved only when the feedback is \textbf{guiding}, \textbf{not forgetful}, \textbf{Precise}, and \textbf{Complete}.
    Overall, the feedback influences the interactions in multiple ways, either negatively or positively, depending on its quality, and also on how LLM can interpret it. 
\end{rqanswer}

\section{Recommendations}
\label{sec:discussion}

We revisit the central question of this paper:
\emph{To what extent can iterative refinement remain effective when the verifier itself is unreliable?}
The TikZ case study suggests a positive but conditional answer.
\textbf{Iterative refinement remains effective under imperfect verification, but only when the verifier is reliable enough to avoid premature acceptance and provides actionable feedback}.
Building on this insight, we derive two sets of recommendations : (i)
practical guidelines for \emph{TikZ customization}, grounded in our empirical results (\citesec{eval_results}), and (ii) broader implications for refinement pipelines relying on \emph{imperfect visual verifiers}.


\subsection{Recommendations for TikZ customization}

\textit{Prefer simple multimodal LLM-based verifiers as a strong baseline.}
As shown in RQ1 (\cref{sec:rq1}), \textbf{Gemini-based verifiers consistently provide the strongest overall assessment results}, whereas more elaborate verification strategies based on decomposition, code generation, or segmentation do not systematically improve either detection quality or alignment with human judgments.
Combined with the results of RQ2 and RQ3, this suggests that the advantage of multimodal verifiers in TikZ is not only that they can assess whether a customization has been applied, but also that they provide feedback that is easier for the generator to exploit during refinement.

\textit{Do not treat these verifiers as reliable or near-perfect oracles.}
Despite strong recall, RQ1 highlights a persistent bias toward false positives.
In an iterative loop, such errors are particularly harmful because they terminate refinement too early.
\textbf{Stopping policies should therefore be conservative, and verifier acceptance treated with caution}.

\textit{Adapt the verifier’s role to the generator’s strength.}
RQ2 shows that \textbf{weaker generators benefit primarily from informative feedback}, whereas \textbf{stronger generators benefit from accurate scoring and early stopping}.
In practice, the verifier should act as a critic for weaker models, and a more selective stopping mechanism for stronger ones. 

\textit{Explicitly optimize for feedback quality.}
RQ3 shows that failures stem not only from incorrect feedback, but also from correct yet ineffective feedback.
Imprecise, non-actionable, incomplete, or forgetful feedback often leads to stagnation, partial application, or incorrect edits.
For TikZ customization, \textbf{effective feedback must localize the problem, specify required changes, remain grounded in the original instruction, and cover all relevant corrections}.

\textit{Assess both requested change and preservation.}
Focusing solely on whether the requested modification is present is not sufficient.
\textbf{Successful customization must preserve the semantics and structure of the original artifact.} 
Verifiers should therefore jointly assess the applied change and the absence of unintended side-effects.

\subsection{Recommendations for imperfect visual verifiers}

Although our empirical claims are carefully restricted to TikZ, the case study suggests broader recommendations for iterative refinement in domains where correctness cannot be checked through deterministic tests.

\textit{Separate decision quality from feedback quality.}
Two distinct capabilities must be considered: (i) the ability to correctly accept or reject outputs, and (ii) the ability to provide guidance that improves them. \textbf{A verifier can be a good scorer but a poor guide, or vice versa}. These dimensions should be evaluated independently.

\textit{Evaluate verifiers within the refinement loop.}
Standalone verification is insufficient. Our results show that \textbf{the same verifier can help or hinder depending on the generator and the interaction dynamics}.
Evaluation should therefore consider final task success, stopping behavior, and the evolution of solutions across refinement steps.

\textit{Prioritize reducing false positives and improving actionability.}
False positives are harmful, as they prematurely terminate refinement.
At the same time, \textbf{useful feedback is not merely correct: it must be precise, guiding, complete, and explicitly tied to the instruction being executed}.
These properties constitute a more realistic target for imperfect visual verifiers than score agreement alone.


\section{Limitations and Threats to Validity}
\label{sec:limitations}

\textit{Case study.}
 Our study focuses on a single language, which limits generalization to other domains. While TikZ customization does not capture the scale and complexity of broader multimodal coding tasks (e.g., web development), it provides, as discussed in \cref{sec:background}, a focused and controlled setting that isolates the core difficulties of the problem while combining code generation, multimodal reasoning, and instruction understanding. Nonetheless, future replication on other languages and settings remains necessary.

\textit{Evaluated models.}
We evaluate only two models, namely Qwen3-VL-30B-A3B-Instruct and Gemini-3-Flash-Preview. This choice is primarily driven by the high annotation cost: even with two models, the study required 8 annotators and 4,289 annotations. The selected models represent different paradigms (\textit{instruct} vs.\ \textit{thinking}), which may limit direct comparability. Nevertheless, both are suitable for the customization task and provide complementary baselines. Importantly, our focus is on imperfect verifiers rather than on the generative models themselves.


\textit{Limitation to 5 steps.}
We restrict evaluation trajectories to five steps. This represents a trade-off between capturing iterative refinement dynamics (e.g., classification quality, trajectory patterns, and feedback quality) and maintaining a manageable annotation effort. Exploring longer trajectories could provide additional insights, but is left for future work.

\textit{Constrained feedback format.}
The system prompts we used enforce a specific feedback format. This limits the diversity of possible feedback. However, as shown in \cref{sec:rq3}, it is sufficient to assess the quality of LLMs' 
feedback in an iterative refinement setting.


\section{Related Work}
\label{sec:related}

Visual code customization sits at the intersection of iterative refinement, critic models, visual question answering, and visual code generation. While each area has seen substantial progress, none directly addresses iterative refinement under an imperfect visual verifier - where feedback may yield false positives, false negatives, or lack the actionability needed to guide meaningful improvement.

\emph{Iterative refinement.}
It was initially introduced as a test-time compute strategy to improve LLM outputs, notably with Self-Refine \cite{madaan_self-refine_2023}, and has since been extended in approaches such as CRITIC \cite{gou_critic_2024}. Prior work has applied iterative refinement to code generation \cite{madaan_self-refine_2023, shinn_reflexion_2023, chen_teaching_2023, zhang_algo_2023, xu_improved_2025} and, more recently, to multimodal tasks \cite{li_metal_2025, xu_improved_2025, azam_multimodal_2024, lee_volcano_2024}. However, these studies do not explicitly consider iterative refinement for structured diagram customization at inference time. Only \citeauthor{xu_improved_2025} and \citeauthor{li_metal_2025} explore both multimodal reasoning and code generation in the context of chart creation, but target full code generation from existing data, rather than customization from existing code. 
In code generation, Post-Hoc iterative refinement typically relies on strong, programmatic feedback signals, similarly to RLVR-based systems. For instance, Self-Debug \cite{chen_teaching_2023} and Reflexion \cite{shinn_reflexion_2023} leverage multi-turn unit test execution to verify intermediate outputs, while Self-Edit \cite{zhang_self-edit_2023} applies a single-turn execution-based feedback strategy. However, these approaches leverage deterministic verification, which is unavailable when correctness depends on properties of rendered diagrams.
Agents also exhibit similar behavior to iterative refinement~\cite{ali_agentic_2025}, by using external tools and verification loops. In contrast to our context, the verifications are usually more reliable.
Other works investigate refinement strategies during training, using feedback loops to improve model parameters \cite{le_coderl_2022, chen_improving_2024, li_relook_2025, liu_agent0-vl_2025}, In contrast, our work focuses exclusively on refinement as a Post-Hoc strategy, applied at inference time.
\emph{Critic models} are LLMs fine-tuned to evaluate or verify outputs \cite{ke_critiquellm_2024, li_generative_2023, kim_prometheus_2024, wang_shepherd_2023, pan_vis-shepherd_2025, li_relook_2025, lee_volcano_2024, qu_ie-critic-r1_2025}. They are often applied in single-turn generate-critique-refine pipelines, sometimes extending to multimodal domains (e.g., \textit{VIS-Shepherd}, \textit{ReLook}, \textit{Volcano}, \textit{IE-Critic-R1}). 
While these critics demonstrate task-specific evaluation, they operate in a single-turn fashion and do not explore multi-step refinement loops. 

\textit{Visual Question Answering and UI critique.} 
The task of Visual Question Answering (VQA) is related to verification in our setting, as verification can be framed as asking: \textit{“Is the instruction \textless instruction\textgreater{} correctly applied to this image?”}. However, unlike standard VQA, our setting requires fine-grained spatial reasoning, instruction grounding, and contextual comparison between an initial and an edited image.
Visual Programming (VP) approaches \cite{suris_vipergpt_2023, ge_recursive_2024, stanic_towards_2024} decompose tasks into sequences of tool calls. While conceptually similar, these methods are primarily designed for natural images and generic visual primitives, limiting their applicability to diagram-level edits.  For completeness, we also evaluate VP-like verifiers(see \cref{base:agentic_vision,sec:base:segtool}).
The task of \textit{UI critique}, which consists of generating design comments from a UI screenshot and design guidelines, is also related. \citeauthor{duan_visual_2025} propose a pipeline that generates feedback and corresponding bounding boxes, and iteratively refines both using a validation model. This approach shares similarities with the segmentation-tool-based verifier (\cref{sec:base:segtool}), but operates on different inputs.

\textit{Visual Code Generation.} 
The intersection of code generation and visual inputs has been explored primarily through image-to-code approaches \cite{zhu_paperbanana_2026, belouadi_detikzify_2024, xu_improved_2025}, and to a lesser extent in editing scenarios \cite{jiang_viscodex_2025, han_chartllama_2023}. However, these approaches focus on single-turn generation, often in a single pass, without any feedback mechanism.
Only \citeauthor{xu_improved_2025} incorporate iterative refinement, but exclusively in the context of image-to-code generation rather than editing.

\textit{Refinement and Repair capability evaluation.} 
Several works evaluate the refinement capabilities of LLMs, either by measuring their ability to improve their own outputs or by assessing the quality of generated feedback. These studies typically focus on domains with deterministic or easily verifiable answers, such as mathematics, algorithms, or multiple-choice reasoning. Notably, \citeauthor{huang_large_2024} analyze self-correction in LLMs and highlight both its potential and its limitations.
 Benchmarks such as RefineBench \cite{lee_refinebench_2025} evaluate refinement using domain-specific checklists and consider both guided (external feedback) and self-refinement settings. Other works, including CriticBench \cite{lin_criticbench_2024}, RealCritic \cite{tang_realcritic_2025}, and CriticEval \cite{lan_criticeval_2024}, focus on evaluating the quality of critics, again in domains where verification can be reduced to clear, often binary outcomes.
Closely related to editing, SR-Eval \cite{zhan_sr-eval_2026} evaluates the ability of LLMs to apply stepwise requirements in software engineering tasks, but considers cumulative sequences of different edits rather than iterative refinement from a single instruction. Similarly, \citeauthor{olausson_is_2024} study self-repair in full code generation settings.
Beyond deterministic domains, \citeauthor{sun_prompt_2024} investigate refinement in text summarization, where correctness is less well-defined, but limit their analysis to single-turn prompting strategies. In the multimodal domain, VF-EVAL \cite{song_vf-eval_2025} evaluates the ability of models to generate feedback on AI-generated videos, but does not consider iterative refinement and is not related to code.

To the best of our knowledge, no existing framework captures the challenges of iterative refinement under imperfect visual verifiers. We 
provide the first novel empirical analysis of multi-step refinement in visual code customization, characterizing verifier behavior with regard to the quality of its classification and feedback.

\vspace*{-2mm}
\section{Conclusion}
In this paper, we explored the applicability and effectiveness of post-hoc iterative refinement with imperfect verifiers, using TikZ code customization as a case study. We introduced an evaluation framework to analyze LLM-based verifiers within an iterative refinement pipeline. Building on the VTikZ dataset, which we refined, we evaluated 20 generator–verifier pairs on 100 TikZ customization tasks, for up to five refinement steps, resulting in a total of 15,345 customized code instances.
 We conducted a large-scale manual evaluation of these customizations, enabling us to analyze both output quality and the impact of feedback in iterative refinement. In particular, we examined verifier performance in (i) assessing whether customization instructions are correctly applied to code, and (ii) understanding how feedback and scoring influence subsequent refinements.

Our results show that verifiers are generally capable of determining whether instructions are applied, but exhibit a bias toward classifying solutions as valid, leading to a high rate of false positives. Feedback improves performance in post-hoc iterative refinement; however, the reliability of scoring becomes more critical for stronger models.
We further derived a study of feedback by systematically analyzing 138 customization trajectories of $\langle$initial customization, feedback, refined customization$\rangle$, extracting fine-grained feedback aspects and their interpretation by LLMs. 
Based on these findings, we provide actionable recommendations not only for TikZ but also for visual verification more broadly. These insights can support the design of more effective and reliable verifiers for further research in TikZ or other similar contexts.

Finally, our results suggest several directions for future work.
The annotated trajectories collected in this study could be used to train a learned scorer for TikZ customization, reducing manual annotation costs and potentially improving the verifier signal available during refinement.
A second direction is to replicate the study on other visual languages, such as SVG or P5.js, to determine which findings transfer beyond TikZ.
A third direction is to investigate imperfect visual verification in richer environments, including web interfaces, games, and 3D scenes, where the artifact to be evaluated is no longer a single rendered image but a more complex interactive environment.
\label{sec:future_work}


\section*{Acknowledgements.} This work is supported by the Inria Défi LLM4Code.

\bibliography{biblio/CriTikz-bib}

\appendix
%

\end{document}